\documentclass[aps,prb,reprint,showpacs,superscriptaddress]{revtex4-1}
\pdfoutput=1 
\usepackage{graphicx} 
\usepackage{amsmath} 
\usepackage{bm} 
\usepackage{natbib}
\usepackage{natmove}
\usepackage{soul}
\usepackage[colorlinks=true, linkcolor=red, citecolor=blue, urlcolor=blue, linktoc=page, bookmarks=false, pdfstartview={FitH}, pdfborder={0 0 0.0 [3 3]}]{hyperref} 
\usepackage{color} 
\usepackage{dcolumn} 
\newcolumntype{.}{D{.}{.}{2.1}}
\newcolumntype{-}{D{.}{.}{4.0}}

\begin{document}

\title{First-principles study of the electronic structure of CdS/ZnSe coupled quantum dots}

\author{Nirmal Ganguli}
\email[Email: ]{NirmalGanguli@gmail.com}
\altaffiliation[Present address: ]{Faculty of Science and Technology and MESA$^+$ Institute for Nanotechnology, University of Twente, P.O.\ Box 217, 7500 AE Enschede, The Netherlands}
\affiliation{Department of Solid State Physics, Indian Association for the Cultivation of Science, Jadavpur, Kolkata 700032, India}
\author{S. Acharya}
\affiliation{Centre for Advanced Materials, Indian Association for the Cultivation of Science, Jadavpur, Kolkata 700032, India}
\author{I. Dasgupta}
\email[Email: ]{sspid@iacs.res.in}
\affiliation{Department of Solid State Physics, Indian Association for the Cultivation of Science, Jadavpur, Kolkata 700032, India}
\affiliation{Centre for Advanced Materials, Indian Association for the Cultivation of Science, Jadavpur, Kolkata 700032, India}

\begin{abstract}
We have studied the electronic structure of CdS/ZnSe coupled quantum dot, a novel heterostructure at the nano-scale. Our calculations reveal CdS/ZnSe coupled quantum dots to be of type-II in nature where the anion-$p$ states play an important role in deciding the band offset for the highest occupied molecular orbitals (HOMO). We show that the offsets of HOMO as well as the  lowest unoccupied molecular orbitals (LUMO) can be tuned by changing the sizes of the components of the coupled quantum dot, thereby providing an additional control parameter to tune the band gap and the optical properties. Our investigations also suggest that formation of alloy near the interface has very little influence on the band offsets, although it affects the spatial localization of the quantum states from the individual components. Comparing the influence of strain on coupled quantum dots and core/shell nanowires, we find strain practically has no role in the electronic structure of coupled quantum dots as the  small effective area of the interface in a coupled quantum dot helps a large part of the structure remain free from any substantial strain. We argue that in contrast to core-shell nanowires, quantum confinement is the key parameter that controls the electronic properties of coupled quantum dot and should therefore be an ideal candidate for the design of a quantum device.
\end{abstract}


\maketitle
\section{Introduction}
Semiconductor heterostructures \cite{loAM11} at nano-scale have attracted considerable attention in the recent times where novel functionalities may be obtained not only by tailoring size and shape of the individual components but also exploiting the combination of the properties of either semiconductors, thereby making their applicability far beyond the limits imposed by the individual nanoparticles \cite{smithACR10}. Modern colloidal techniques allow fabrication of various types of heterostructures such as core shell nanocrystals (NC) \cite{kimJACS03}, multicomponent hetero-nanorods \cite{KumarSmall07}, tetrapods \cite{MillironN04} and very recently heterodimers \cite{TeranishiJPCL13} and coupled quantum dots \cite{senguptaAM11}. Semiconductor heterostructures are typically classified either as type-I or type-II, depending on the relative alignment of the conduction and the valence band edges of the materials that constitute the interface. In a type-I heterostructure, the alignment of the bands is such that both conduction and valence band edges of semiconductor~A (smaller band gap) are located within the energy gap of semiconductor~B (larger band gap), so that the electron and hole pairs excited near the interface tend to localize in semiconductor~A.
For a type-II heterostructure, the relative alignment of the conduction and valence bands of the constituent materials offer a spatially indirect band gap resulting in an optical transition energy smaller than the band gap of either of the constituent materials. As a consequence of this staggered alignment of bands, the lowest energy states for the electrons and the holes are in different semiconductors which is highly attractive for applications in photovoltaics, where such charge separation is desirable. \cite{loAM11, ivanovJACS07, MillironN04}

Tuning the optical properties of semiconducting nano heterostructures can be achieved by selection of the constituent materials and taking  advantage of additional parameters such as size dependent quantum confinement exhibited by the systems at nanometer scale. In addition, type-II heterostructures offer an attractive possibility of controlling the effective band gap by engineering the band offsets at the interface \cite{senguptaAM11}. Another parameter that has profound impact on the electronic structure and band offsets in nano heterostructures is the strain resulting due to sharp lattice mismatch of the constituents at the interface \cite{SmithNN09, ramprasadJPCC10}. While nano-scale heterojunctions can tolerate larger lattice mismatch in comparison to its bulk counterpart, the resulting strain may further shift energy levels and band offsets in a non-trivial way \cite{yangNL10}. It has been shown that the strain induced change in the band gap may be comparable to that induced by quantum confinement in highly lattice mismatched nano-scale heterojunctions \cite{yangNL10}. Recently it has been illustrated that strain can be advantageous in tuning the optical properties of core-shell nanocrystals \cite{SmithNN09}. Epitaxial deposition of a compressive shell (ZnS, ZnSe, ZnTe, CdS or CdSe) onto a nanocrystalline core (CdTe) produces strain that changes standard type-I band alignment to type-II behavior, ideal for application in photovoltaics \cite{SmithNN09}. On the other hand in some cases strains produced at the interface may be relieved by creating dislocations at the interface giving rise to non-radiative decay channels proving to be highly detrimental for applications \cite{SarmaJPCL10}.
In this respect, recently suggested type-II nano-heterostructures obtained by coupling semiconductor quantum dots are interesting as they possibly rely on controlling the effective transition energy gap by engineering the band-offsets at the interface primarily by quantum confinement as the effect of strain in such systems is expected to be small. This is due to the fact that the actual area of the interface is much smaller in coupled quantum dots in comparison to core-shell nano-systems due to their difference in geometry, which substantially reduces the stress in case of the former. As the effect of strain is expected to be minimal, quantum confinement is of prime importance in coupled quantum dots providing an ideal opportunity to design interface as a quantum device that may find application either in optoelectronic devices(e.g. photovoltaic device) or for the realization of qubits for quantum information processing.\cite{quantuminfo} It is interesting to note that a recent report on coupled semiconductor quantum dots of CdS/ZnSe demonstrated the tuning of photoluminescence wavelength (a manifestation of the effective gap) over a large range of $\sim$100~nm simply by changing the ratio of the component sizes constituting the coupled quantum dot\cite{senguptaAM11} that clearly demonstrates the tunability of optical properties via controlling  band-offsets primarily due to quantum confinement.

The electronic structure at the interface of nano-heterostructures plays a crucial role in tailoring the band gap and band offsets.
In the present paper using density functional theory in the framework of generalized gradient approximation (GGA) we have investigated in details the electronic structure of coupled quantum dots and compared them with core shell nanowires. While a well known limitation of  GGA is that it tends to underestimate the band gaps and does not provide reliable estimate of band offsets between chemically dissimilar materials \cite{StevanovicPCCP14, *GruneisPRL14}. However, this is not a matter of concern in this paper as the materials considered here (CdS, CdSe, ZnSe) exhibit very similar quasiparticle shifts \cite{StevanovicPCCP14, *GruneisPRL14} so the value of the offsets may not change significantly. Further we shall discuss the trends and the physical origin of band offsets which are not dependent on the actual value of the band gap and band offsets. We have studied in details the coupled quantum dots of CdS and ZnSe from first principles density functional calculations to understand the nature of chemical interaction at the interface, that leads to the type-II nature of the heterojunction. We have also calculated the band offsets and investigated how it changes with variation of the component size. As the role of strain remained unexplored in coupled quantum dots, we have therefore calculated the strain profile and the impact of strain on band offsets and compared our results with CdS$_\text{core}$/ZnSe$_\text{shell}$ nanowires. In the following we shall argue that band offsets in coupled quantum dots are primarily dictated by the interaction between the anion-$p$ states along with quantum confinement, making them ideal for a quantum device.

\section{Computational detail and Simulated Structure}
All the electronic structure calculations presented here are performed using {\em ab~initio} density functional theory (DFT) as implemented in Vienna ab-initio simulation package (VASP) \cite{vasp1, *vasp2}. Projector augmented wave (PAW) method\cite{paw} along with plane wave basis set are used for our calculations. PAW potentials with 12 valence electrons (4d$^{10}$ 5s$^{2}$) for Cd, 12 valence electrons (3d$^{10}$ 4s$^{2}$) for Zn, 6 valence electrons (3s$^{2}$, 3p$^{4}$) for S and 6 valence electrons (4s$^{2}$, 4p$^{4}$) for Se with an energy cut-off of 500 eV for the plane wave expansion of the PAW's was employed in our calculations. The exchange-correlation (XC) part is approximated through generalized gradient approximation (GGA) due to Perdew-Burke-Ernzerhof (PBE)\cite{pbe}. The dangling bonds at the surface of the clusters as well as the nanowires are saturated using fictitious hydrogen atoms with fractional charges, as proposed by Huang~{\em et~al.}\cite{chelikowskyPRB05}. Our calculations are performed in the framework of periodic boundary condition and the periodic images of the clusters and the nanowires along the transverse directions are separated by vacuum layers of sufficient width ($\sim$10~\AA). In view of the large size of the simulation cell (tiny Brillouin zone) we have employed only one k-point ($\Gamma$ point) for the coupled quantum dots and a $\Gamma$ centered $k$-mesh of 1$\times$1$\times$8 for the nano-wires. The atomic positions were relaxed to minimize the Hellman-Feynman force on each atom with a tolerance of 0.01~eV/\AA. The lattice strain in the heterostructures were calculated following the method proposed by Pryor{\em ~et~al.}\cite{pryorJAP98}

\section{\label{sec:result}Results and discussions}
\subsection{Coupled quantum dots of similar size}
\begin{figure}[tb]
        \centering
        \includegraphics[scale = 0.27]{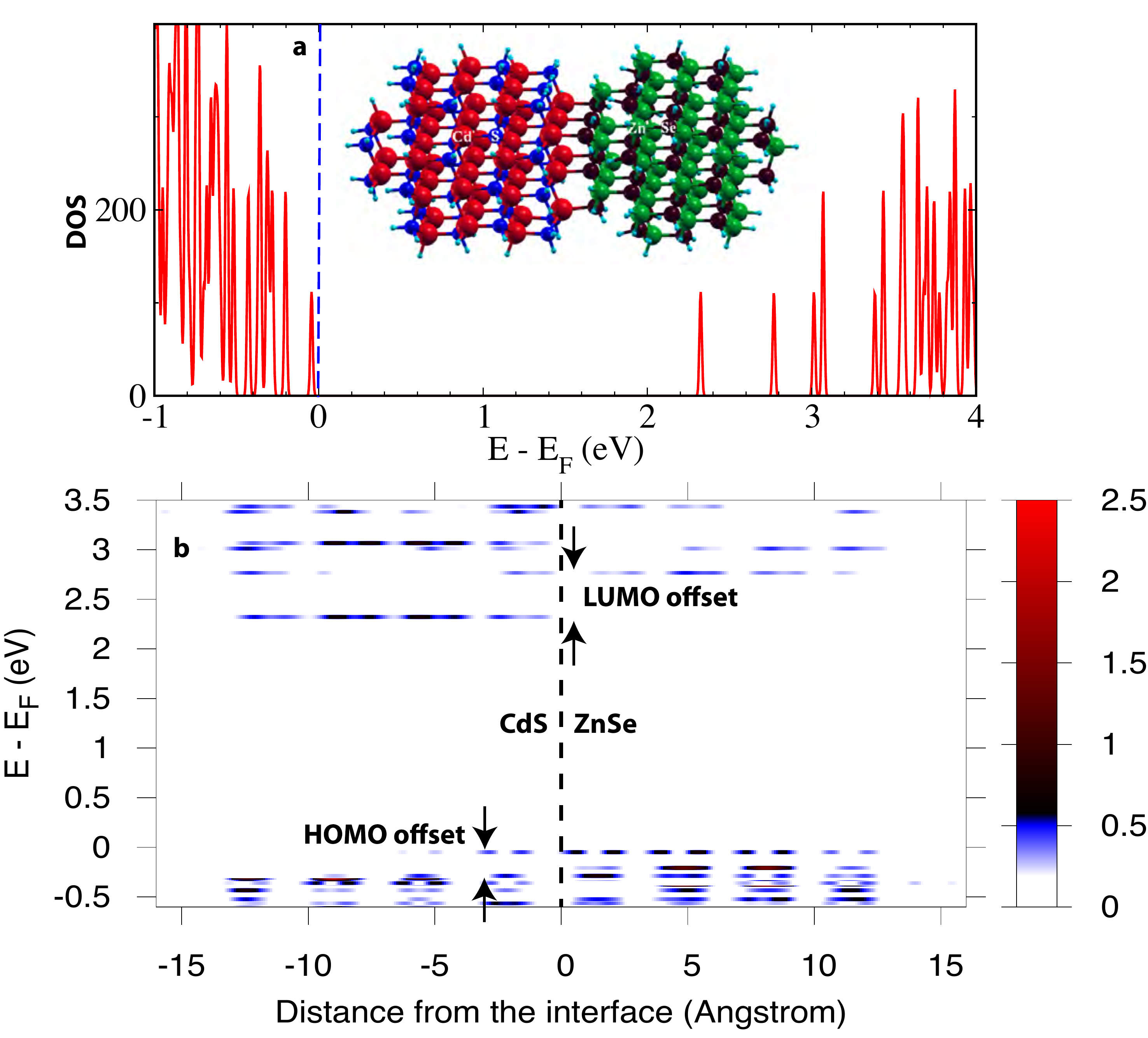}
        \caption{\label{fig1} (Color online) (a) The total density of states for coupled CdS/ZnSe quantum dots of similar size. The inset shows the structure of the coupled quantum dots, where red, blue, green, maroon, and light blue balls indicate Cd, S, Zn, Se, and the fictitious passivator atoms respectively (this convention has been followed throughout the article). (b) The energy resolved charge density has been plotted as a function of the distance from the interface.}
\end{figure}

To begin with, we have simulated the coupled quantum dot formed by coupling of CdS and ZnSe quantum dots of similar sizes. Following the experimental observations,\cite{senguptaAM11} we have taken the CdS cluster in the wurtzite (hexagonal) phase and the ZnSe cluster in the zinc blende (cubic) phase. In our simulation, the CdS cluster consists of 45 Cd atoms and 51 S atoms, whereas the ZnSe cluster comprises 44 Zn and 46 Se atoms. The diameter of both the clusters are $\sim 1.6$~nm. The heterostructure is formed by attaching the polar (0001) facet of CdS cluster in wurtzite structure with the polar (111) facet of ZnSe cluster in cubic structure,  where the Cd-terminated polar facet of the CdS cluster binds to the Se-terminated polar facet of the ZnSe dot as shown in the inset of Figure~\ref{fig1}(a).
The total density of states (DOS) corresponding to the coupled quantum dot of CdS-ZnSe is shown in Figure~\ref{fig1}(a). The DOS suggest that the gap between the highest occupied molecular orbital (HOMO) and the lowest unoccupied molecular orbital (LUMO) for the coupled quantum dot is 2.35~eV. This value of the gap is smaller than the calculated  gaps for both of its components namely the CdS cluster ($\sim$ 2.65~eV) and  that of the ZnSe cluster ($\sim$ 3.10~eV). The trend in the calculated gap of the components is consistent with the experimental band gap of bulk CdS (2.42~eV) and ZnSe (2.70~eV). The calculated gaps for the individual dots are larger compared to the bulk experimental values due to quantum confinement but are possibly underestimated due to the usual limitation of GGA. The effective gap of the heterostructure being less than either of the components indicates that the band alignment at the interface may be of type-II. 

To obtain further insights on the nature of the band alignment, the character of the HOMO and the LUMO states and to estimate the offsets for HOMO and LUMO at the interface we have calculated the energy resolved charge density along the direction perpendicular to the interface. In order to evaluate the energy resolved charge density, the band decomposed charge density corresponding to a particular energy eigenvalue is calculated for each $k$-point. The resulting charge density is averaged over planes parallel to the interface. This averaged charge density for a given energy at a particular $k$-point scaled by an arbitrary constant (same constant is used for all calculations) is plotted as a function of the distance from the interface. Such spatially averaged charge densities reflect the spatial distribution of every state perpendicular to the interface within a suitable range of energy. This energy resolved charge density is particularly useful for visualizing band alignment of nano-scale heterostructures where either one or only few k-points are used for the calculation. The energy resolved charge density for CdS/ZnSe coupled quantum dot is  shown in Figure~\ref{fig1}(b). We gather from the figure that the highest occupied state is primarily confined to the ZnSe part of the coupled quantum dot  with a short tail extended to the CdS part. On the other hand, the lowest unoccupied state is confined to the CdS part. Further, this figure offers us a clear view of the localization of all the states in the energy range of interest. From this figure, we have obtained the values of the offsets for HOMO and LUMO to be 0.27 and 0.45~eV respectively. In addition, we have also calculated the valence band offset following the method suggested by \citet{HinumaPRB13}. In this approach an electrostatic potential averaged within a PAW sphere at an atomic site is taken as the reference level for the evaluation of the valence band offsets. The calculated valence band offsets using this method are shown in Table~\ref{tab:offset}. We find both the method provide nearly identical value for the HOMO offsets. The LUMO offsets calculated using the latter method, {\em i.e.} adding the HOMO offsets and the difference of energy gaps of the individual pristine clusters, show similar trend as obtained from the energy resolved charge density method but have substantially larger values. The reason for this discrepancy may be attributed to the fact that the energy gaps of CdS and ZnSe obtained from the CdS and ZnSe part of the coupled quantum dot are smaller than their values in the respective pristine clusters due to the mitigation of quantum confinement. Upon formation of the heterostructure, we find the energy gap of the ZnSe part to become substantially smaller, while the gap of CdS part reduces marginally. This effect is not accounted for within the latter method for calculating the LUMO offsets, leading to a systematic overestimation of the same.

\begin{table*}
	\caption{\label{tab:offset} The band offsets (in eV) for coupled quantum dots calculated (i) using energy resolved charge density plots and (ii) using average electrostatic potential at the atomic PAW spheres as suggested by \citet{HinumaPRB13}, are tabulated here.}
	\begin{ruledtabular}
		\begin{tabular}{lcccc}
			Heterostructure & \multicolumn{2}{c}{Charge density method} & \multicolumn{2}{c}{Electrostatic potential method} \\
			\cline{2-3}  \cline{4-5}
			 & HOMO & LUMO & HOMO & LUMO \\
			 & offset & offset & offset & offset \\
			\hline
			CdS/ZnSe 1:1 & 0.27 & 0.45 & 0.29 & 0.74 \\
			CdSe/ZnSe    & 0.00 & 0.64 & 0.06 & 0.96 \\
			CdS/ZnSe 2:1 & 0.30 & 0.94 & 0.37 & 1.20 \\
			CdS/ZnSe diffused (1$^\text{st}$ bilayer) & 0.26 & 0.46 & 0.34 & 0.79 \\
			CdS/ZnSe diffused (2$^\text{nd}$ bilayer) & 0.21 & 0.42 & 0.34 & 0.79 \\
		\end{tabular}
	\end{ruledtabular}
\end{table*}

\begin{figure}[tb]
        \centering
        \includegraphics[scale = 0.33]{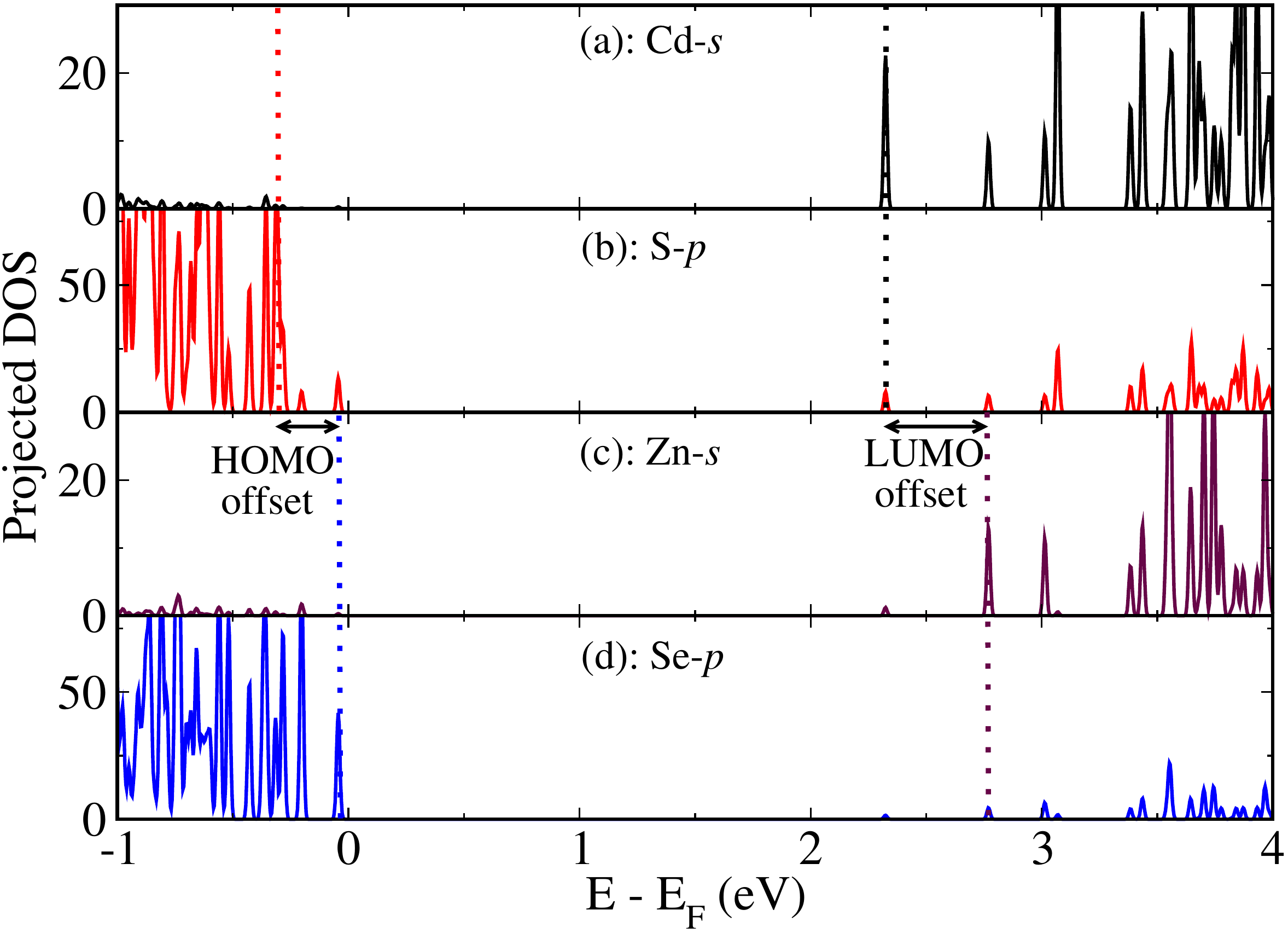}
        \caption{\label{fig2} (Color online) The density of states projected onto (a): Cd-$s$, (b): S-$p$, (c): Zn-$s$, and (d): Se-$p$ states for coupled CdS/ZnSe quantum dots of similar size have been shown here.}
\end{figure}

We have next investigated the role of anion-$p$ states on the valence band offset. We have plotted the partial DOS for the anion-$p$ states and cation-$s$ states (see Figure~\ref{fig2}) to understand the interaction at the interface of the heterojunction that leads to the offset.
The partial DOS indicates that the occupied states near the gap are primarily anion-$p$ like which also has some admixture with the cation-$d$ states, whereas the unoccupied states near the gap mainly show cation-$s$ like character, with a little admixture with anion-$p$ states. Notably, the offsets between the $p$ states of S and Se, and the $s$ states of Cd and Zn turn out to be the same as the calculated HOMO and LUMO offsets respectively. This observation corroborates the picture  that the interaction between the anion-$p$ states admixed with cation-$d$ states are crucial for the  HOMO offset \cite{kroemerPRB77} with only implicit role of the  cation-$d$ states. The coupling between the anion-$p$ and cation-$d$ states however play an important role for bulk semiconductor heterostructures as suggested by \citet{weiAPL98}.

To ascertain further the role of anion-$p$  and cation-$d$ states in determining the HOMO (valence band) offsets, we have simulated a similar  heterostructure comprising CdSe and ZnSe quantum dots: {\em i.e.} the same anion for both the components. The CdSe/ZnSe heterostructure is simulated by replacing all the sulphur atoms of CdS/ZnSe quantum dots heterostructure with selenium atoms followed by optimization of the atomic positions. 
Our calculated band gap (2.20~eV) for CdSe cluster is found to be smaller in comparison to the  ZnSe cluster (3.10~eV) and is consistent with the experimental bulk band gap of CdSe (1.73~eV) and ZnSe (2.70~eV). The effective gap is found to reduce marginally (by 0.14 eV) upon formation of the heterojunction. The band offsets calculated using the energy resolved charge density is listed in Table~\ref{tab:offset} and shows no offset for HOMO and a large offset for LUMO. Only a small value for the HOMO offset is obtained from the method of \citet{HinumaPRB13} (See Table~\ref{tab:offset}). Hence the reduction in effective HOMO-LUMO gap upon formation of heterojunction may be attributed to the increase in the system size and thereby reduction of the band gap due to mitigation of quantum confinement. The quasi type-II nature of the heterojunction (i.e.\ no offset for HOMO and substantial offset for LUMO) supports the common-anion rule, \cite{kroemerPRB77} where it is argued that no offset for valence band should be found for heterojunctions with common anion for both the components confirming the important role played by the anion-$p$ states in determining the valence band offset.

\subsection{Variation of component size}
Having confirmed that an ideal CdS-ZnSe coupled dot lead to type-II heterostructure, next we have explored the effect of variation of component size on the offsets of HOMO and LUMO. In this context, recent experiments suggest that photoluminescence wavelength increase with increasing concentration of CdS ({\em i.e.\ } increasing size of CdS quantum dot),\cite{senguptaAM11} which may be attributed to the change in band offset. Engineering the band offset by modifying the sizes of the components in a coupled quantum dot is an attractive feature  that may find application for device fabrication. In view of the above, we have simulated a coupled quantum dot where the number of atoms in CdS cluster is nearly double the number of atoms in ZnSe cluster.

\begin{figure}[tb]
        \centering
        \includegraphics[scale = 0.32]{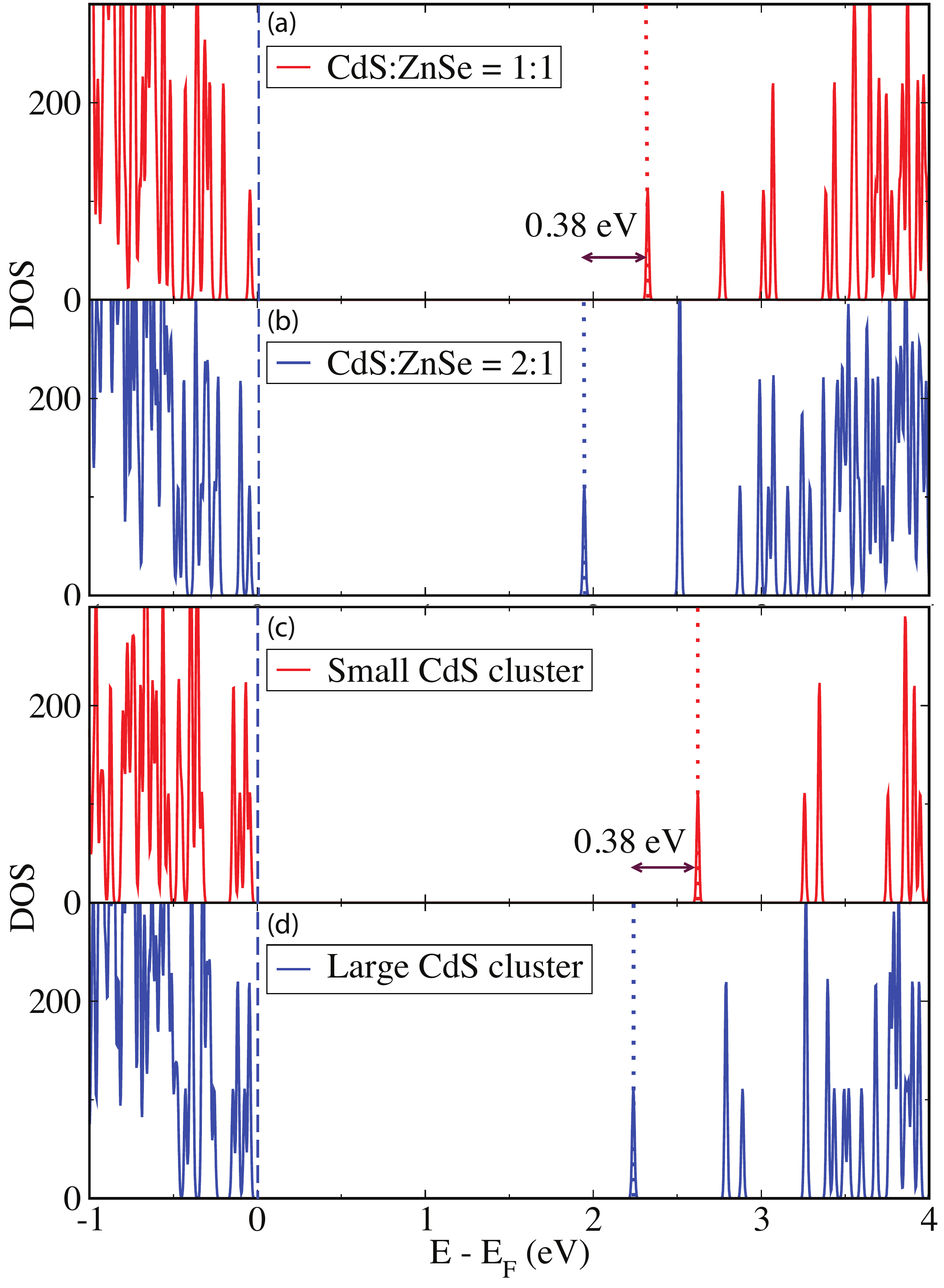}
        \caption{\label{fig4} (Color online) This figure compares the density of states for CdS/ZnSe 1:1 dots (a) to CdS/ZnSe 2:1 dots (b) {\em vis-\`{a}-vis} the density of states for the small and the large CdS quantum dots (c,d).}
\end{figure}

We have therefore considered a heterojunction of CdS/ZnSe clusters where a CdS cluster consisting of 93 Cd atoms and 96 S atoms is coupled  to a  ZnSe cluster consisting of 44 Zn and 46 Se atoms. We shall refer to this system as 2:1 system, whereas the system with similar component sizes studied earlier will be referred to as 1:1 system. The densities of states for the 1:1 system and the 2:1 system is displayed in Figure~\ref{fig4}(a) and Figure~\ref{fig4}(b), respectively, a comparison between them shows that the effective gap between HOMO and LUMO decreases by 0.38~eV upon increasing the size of CdS cluster.
It is interesting to note that the gap between HOMO and LUMO for pristine CdS cluster also decreases by the same amount (0.38~eV) upon increasing the size, as seen from Figure~\ref{fig4}(c) and Figure~\ref{fig4}(d). The reduction of the  effective gap for the heterojunction is a result of mitigation of quantum confinement due to the large size of the CdS cluster. 
The band-offsets calculated using two different method are listed in Table~\ref{tab:offset}.
Hence increasing the size of one of the components (here CdS) of a coupled quantum dot reduces the HOMO-LUMO gap for that component due to quantum confinement that primarily changes the LUMO offset while the offset between the HOMO states does not change much.
The above discussion points to the fact that variation in the size of the components for a coupled quantum dot heterojunction is a novel control parameter that provides an opportunity to tune the offsets for suitable applications. More importantly, the range of effective gap thereby accessible may be far beyond the range accessible by manipulating the size of an individual quantum dot.

\subsection{Diffused interface}
\begin{figure}
        \centering
        \includegraphics[scale = 0.27]{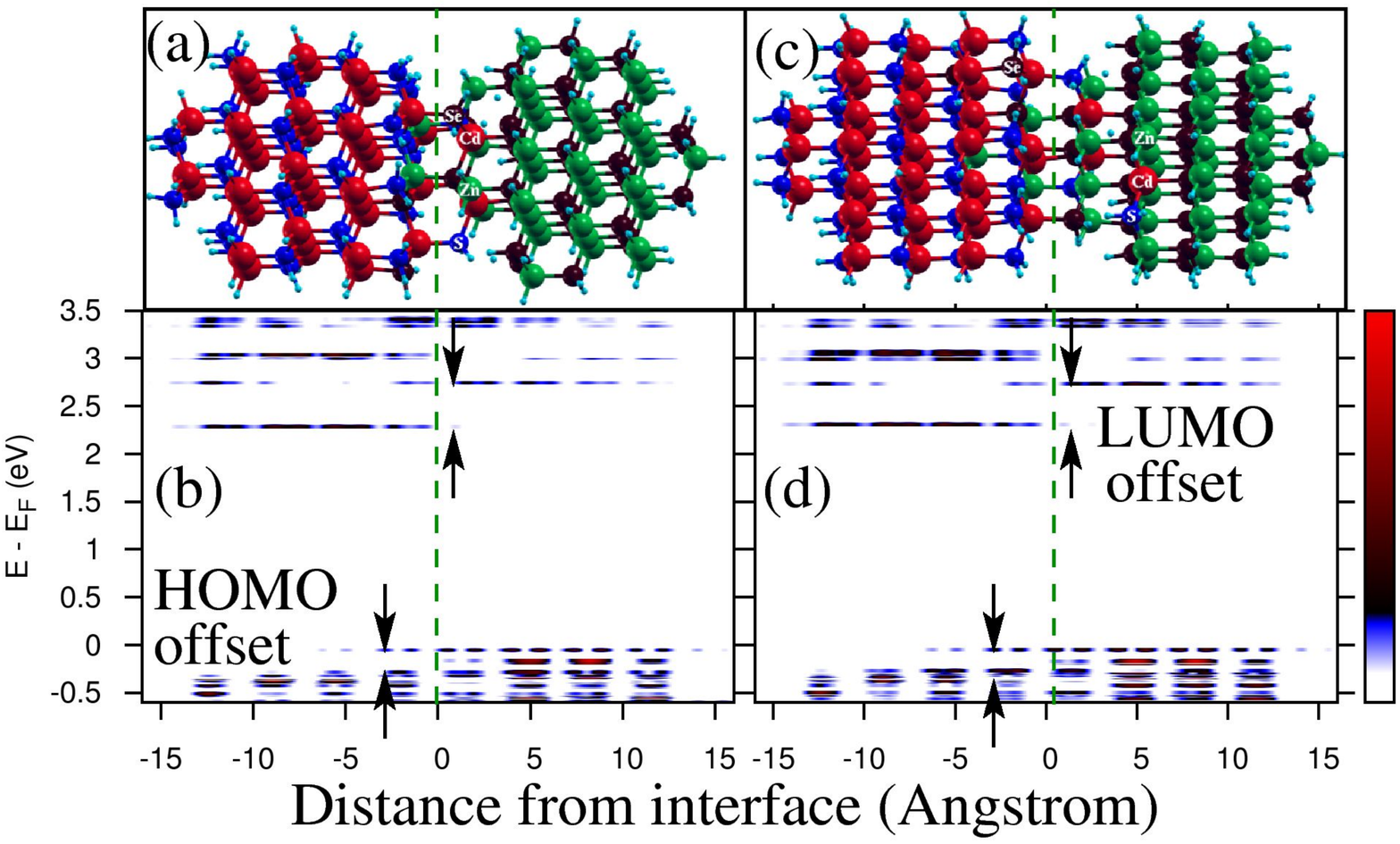}
        \caption{\label{fig6} (Color online) (a) and (b) [(c) and (d)] show the structure and the energy resolved charge density respectively for the diffused (1$^\text{st}$ bilayer) [diffused (2$^\text{nd}$ bilayer)] system.}
\end{figure}

In the preceding discussion we have considered ideal interface but it is quite likely that the interface of the coupled dots may be a diffused alloy of CdS and ZnSe. In view of the above, we have examined the influence of interlayer diffusion of the atoms near the interface by simulating two different alloyed heterostructures, namely: (i) where two Cd (S) atoms replace two Zn (Se) atoms and vice versa at the first cationic (anionic) interlayer, {\em i.e.} the diffusion is restricted to the first bilayers near the interface, (see Figure~\ref{fig6}(a)) and (ii) where in addition to (i) one Cd (S) atom replaces one Zn (Se) atoms and vice versa at the second cationic (anionic) interlayer, {\em i.e.} the diffusion extends upto the second bilayers near the interface (see Figure~\ref{fig6}(c)). The energy resolved charge density plot for these  diffused interfaces are shown in Figure~\ref{fig6}(b) and Figure~\ref{fig6}(d) respectively. The calculated band offsets are listed in Table~\ref{tab:offset}.
These values compare well with the ideal 1:1 interface. Our observations indicate that the offsets are not very sensitive to the diffusion at the interface as the inter-layer diffusion possibly does not influence the interaction between anion-$p$ and cation-$s$ states significantly but it affects the spatial localization of the states that may be detrimental for carrier separation required for photovoltaic applications. We do not observe localized interface induced states for perfectly passivated coupled quantum dots. However, lack of fictitious H atoms near the interface may lead to such localized (dangling bond) states.  Coupled dots prepared using colloidal technique usually have long chain organic molecules passivating  the dangling bonds.

\subsection{\label{sub:nanowire}CdS$_\text{core}$/ZnSe$_\text{shell}$ nanowire heterojunction}
\begin{figure}
        \centering
        \includegraphics[scale = 0.26]{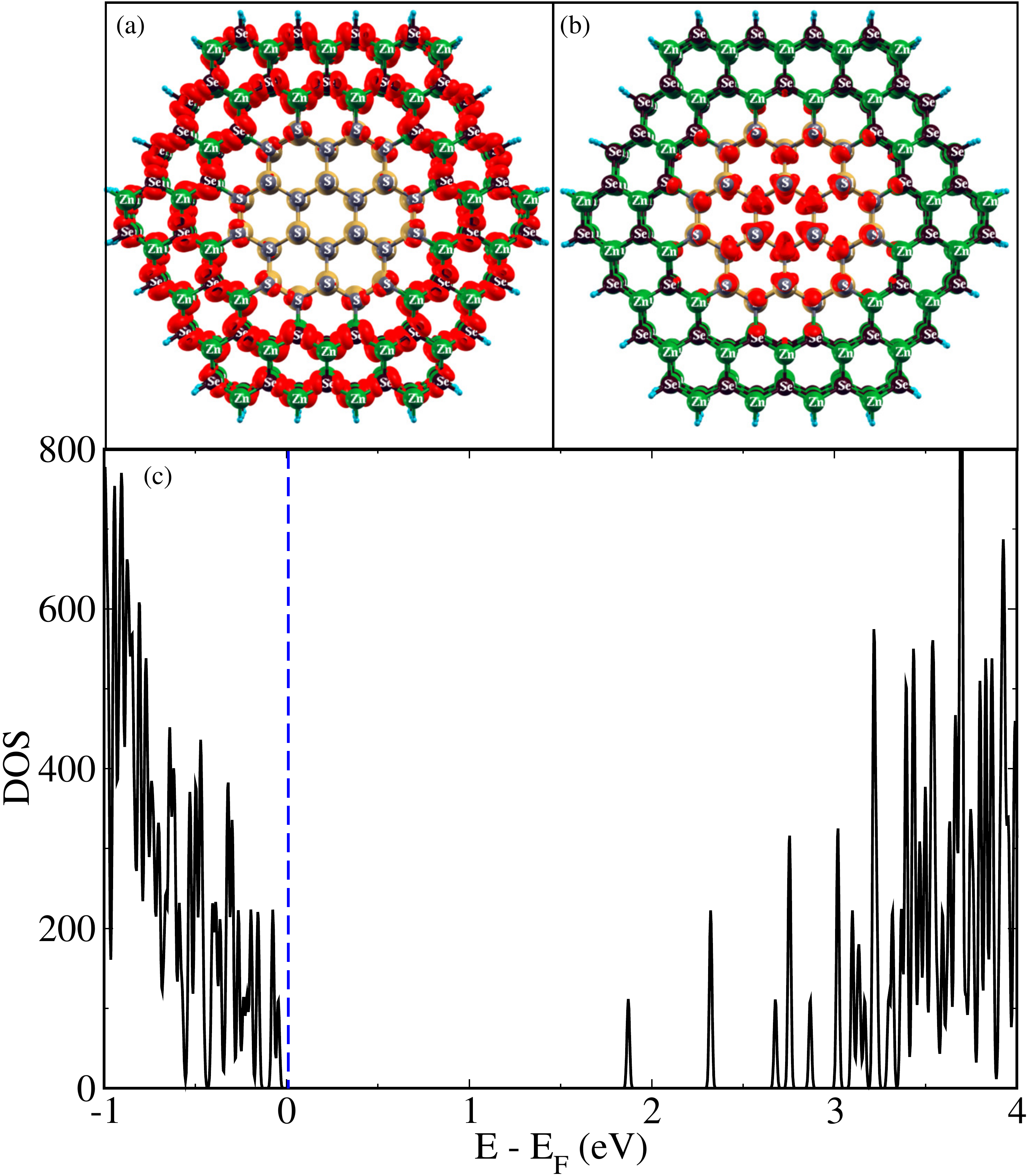}
        \caption{\label{fig7} (Color online) The charge density isosurfaces corresponding to HOMO and LUMO of the CdS$_\text{core}$/ZnSe$_\text{shell}$ nanowire heterojunction have been shown in (a) and (b) respectively. (c) displays the total density of states for the nanowire heterostructure.}
\end{figure}

After investigating the coupled quantum dot heterojunctions in details, we have studied CdS$_\text{core}$/ZnSe$_\text{shell}$ nanowire heterojunctions, where we anticipate significant difference in electronic structure because of larger interfacial area at the interface. In order to simulate the heterojunction we have assumed two rings of CdS in the wurtzite structure as core, surrounded by two rings of ZnSe in the wurtzite structure as shell, as shown in the cross sectional view of the nanowire (see Figure\ref{fig7}(a-b)).
The radius of this cylindrical heterostructure is $\sim$13~\AA. The dangling bonds at the surface are properly saturated by fictitious hydrogen atoms with fractional charge \cite{chelikowskyPRB05}. The crystallographic $c$ direction of wurtzite structure has been assumed to be the growth direction of the nanowire. Unlike coupled quantum dots, here the interface is not formed by attaching polar facets.

The density of states for the CdS$_\text{core}$/ZnSe$_\text{shell}$ nanowire heterojunction is shown in Figure~\ref{fig7}(c). We note that the effective band gap for this system is calculated to be 1.90~eV. The smaller value of the effective band gap for the core-shell nanowire may be attributed to the absence of confinement along the $c$ direction that reduces the band gaps for both the components constituting the nanowire.

\begin{figure}
        \centering
        \includegraphics[scale = 0.28]{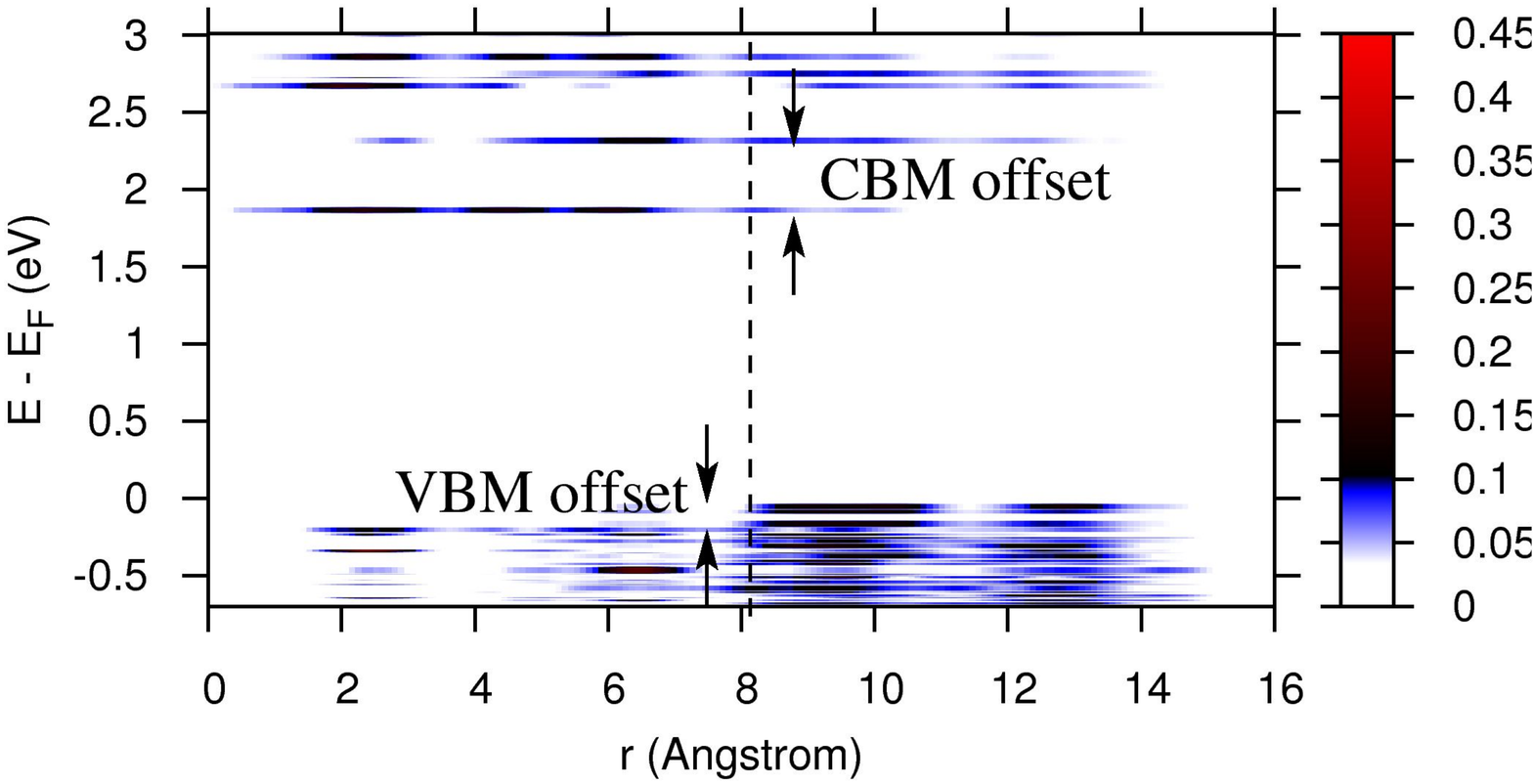}
        \caption{\label{fig8} (Color online) The energy resolved charge density as a function of the radial coordinate of the cylindrical nanowire heterojunction has been depicted here.}
\end{figure}

The charge densities corresponding to valence band maximum (VBM) and conduction band minimum (CBM) are shown in Figure~\ref{fig7}(a) and Figure~\ref{fig7}(b) respectively. From this figure we find the VBM and the CBM to be confined in the shell and the core region respectively, confirming type-II nature of the heterojunction. The energy resolved charge density plot for this system is displayed in Figure~\ref{fig8},
which indicates the VBM and the CBM offsets to be 0.20~eV and 0.44~eV respectively. In comparison to the offsets calculated for the coupled quantum dots, the VBM offset is smaller in this case while the CBM offset is nearly the same.
\subsection{\label{sub:strain}Effect of strain}
\begin{figure*}
        \centering
        \includegraphics[scale = 0.3]{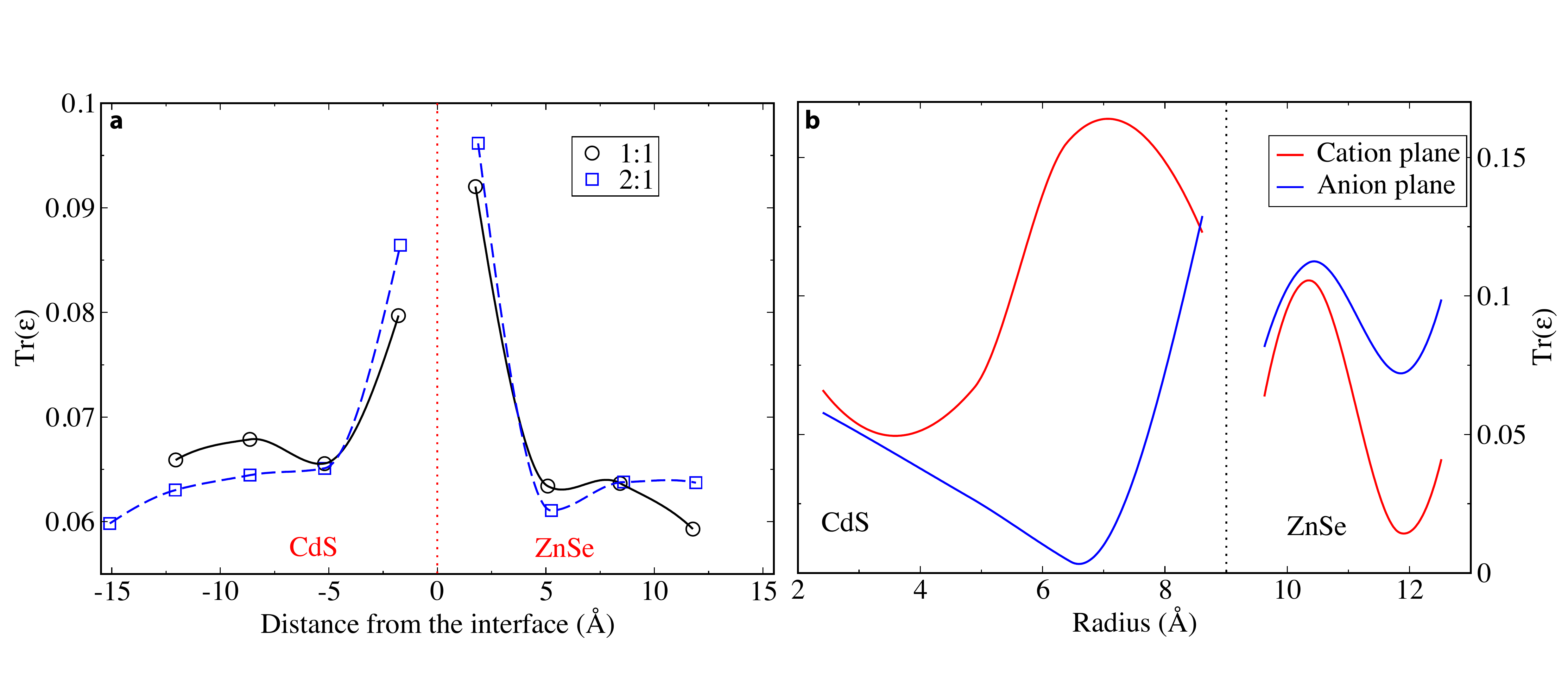}
        \caption{\label{fig9} (Color online) The trace of the strain tensor has been plotted as a function of distance along the perpendicular direction to the interface for (a) the coupled quantum dots with varying size and (b) the CdS$_\text{core}$/ZnSe$_\text{shell}$ nanowire.}
\end{figure*}

As discussed earlier, the band alignment of nanoscale heterojunctions should be very sensitive to the lattice strain. In terms of the effect of strain, the coupled quantum dots and core/shell nanowires are expected to be very different. We have calculated the strain profiles for coupled quantum dots and core/shell nanowires by using an atomistic model \cite{pryorJAP98} for elasticity where the parameters of the model are calculated using  {\em ab~initio} electronic structure calculations within density functional theory.
We have calculated the trace of the strain tensor that represents the volumetric strain for the system. Our results for the volumetric strain for coupled dot (for 1:1 and 2:1) and core-shell nanowrires are displayed in Figure~\ref{fig9}(a) and Figure~\ref{fig9}(b) respectively.
We note from Figure~\ref{fig9}(a) that the strain profiles for coupled quantum dots do not change significantly with variation of component size. As expected, the strain is quite large near the interface and it sharply decreases as we move away from the interface. On the other hand, for the core/shell structure the strain profiles are shown for the cationic and the anionic planes separately in Figure~\ref{fig9}(b) and we find an oscillatory nature of the strain field where the value of the strain may be significantly large even far away from the interface. Comparing the strain profiles for both the systems we gather that the core/shell nanowire is more strained compared to the coupled quantum dots due to the large interface of the latter.

\begin{figure*}
        \centering
        \includegraphics[scale = 0.5]{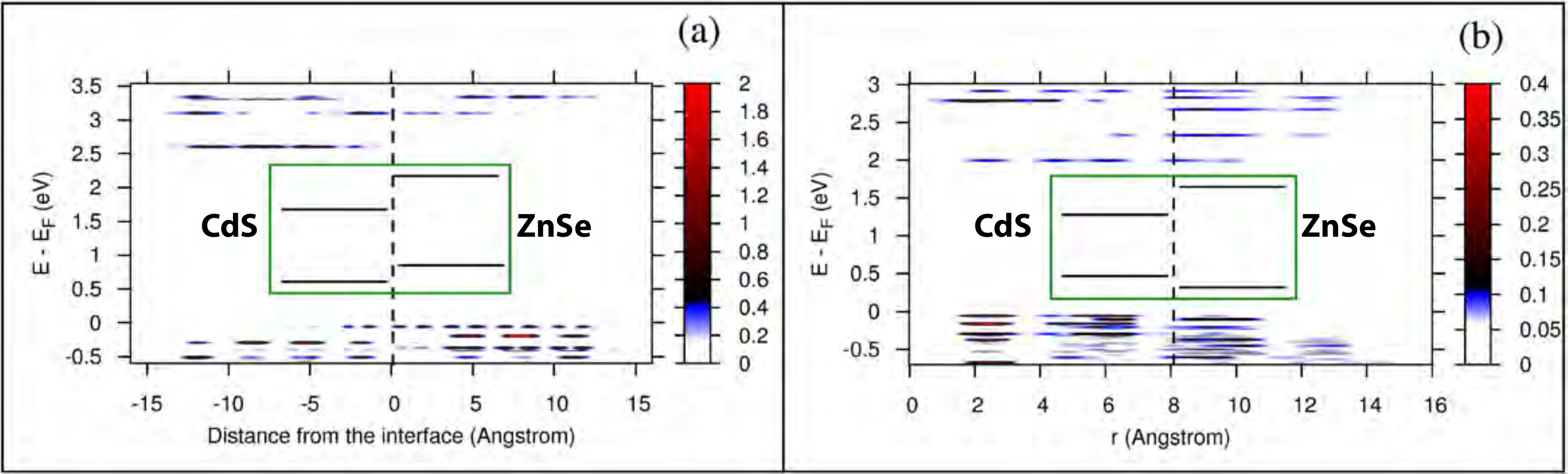}
        \caption{\label{fig10} (Color online) The energy resolved charge density plots for unrelaxed systems: (a): CdS/ZnSe coupled quantum dots (HOMO offset: 0.29 eV, LUMO offset: 0.47 eV, type-II) and (b): CdS$_\text{core}$/ZnSe$_\text{shell}$ nanowire (VBM offset: 0.09 eV, CBM offset: 0.33 eV, type-I). Insets show the schematic band alignments.}
\end{figure*}

In order to quantify the effect of strain on the alignment of bands, we have calculated the band offsets for the unrelaxed (discretely strained at the interface) heterostructures of CdS/ZnSe in a coupled dot as well as core/shell nanowire geometry. A similar model was employed earlier to study the impact of strain on band gaps in core-shell nanostructures. \cite{yangNL10} The model for unrelaxed coupled quantum dots comprise a CdS and a ZnSe quantum dot with bond lengths and the bond angles matching the corresponding bulk structures in wurtzite and zincblende forms respectively. The heterojunction is formed by bringing Cd-terminated (0001) plane of CdS close to Se terminated (111) plane of ZnSe. The separation between Cd and Se planes is 2.44~\AA. The dangling bonds are passivated by fictitious hydrogen atoms with fractional charge, located at a distance of 1.25~\AA. On the other hand, the unrelaxed CdS$_{\text{core}}$/ZnSe$_{\text{shell}}$ nanowire heterostructure consists of both of its components in wurtzite form with their respective bulk bond length and bond angle values. The dangling bonds at the ZnSe$_{\text{shell}}$ part are passivated by fictitious hydrogen atoms with fractional charge, located at a distance of 1.25~\AA.
The corresponding energy resolved charge density plots for the unrelaxed structures are shown in Figure~\ref{fig10}.
The band-offsets for the un-relaxed 1:1 coupled quantum dots (HOMO offset: 0.29 eV, LUMO offset: 0.47 eV, type-II) are nearly identical to that obtained for the relaxed coupled quantum dots (HOMO offset: 0.27 eV and LUMO offset = 0.45 eV) indicating the effect of relaxation ({\em i.e.} distribution of strain over the structure) is negligible for coupled quantum dots. On the other hand the impact of the distribution of strain is appreciable in the core-shell nanowire.  In the core-shell nanowire, upon relaxation not only the value of the band offset change appreciably but also the band alignment become type-II, (see Figure~\ref{fig8}) for the relaxed system whereas unrelaxed system (see Figure~\ref{fig10}(b)) shows type-I nature of alignment. The band-offset for the unrelaxed (relaxed) core-shell nanowire is calculated to be VBM offset : 0.09eV (0.20eV) and CBM offset: 0.33 eV (0.44eV) where both the VBM and CBM offsets of the unrelaxed structure is substantially different from the relaxed structure.  From the quantitative estimate of the volumetric strain and the comparison of the effect of strain on the band-offsets for the coupled quantum dot and core-shell nanowire  we understand that as opposed to coupled quantum dots, the electronic structure is very sensitive to strain for core/shell nanowires due to the large interfacial area of the latter. While strain is an important factor in determining the band-offsets in core-shell nanowires, quantum confinement is the only key deciding factor for band-offsets in coupled quantum dots.

\section{\label{sec:conc}Conclusion}
In conclusion, we have studied the electronic structure of coupled quantum dots consisting of CdS/ZnSe clusters in details to understand the origin and nature of band offset. We have also explored in details the impact of the variation in component size and lattice strain on band offset of the coupled quantum dots. We have found  the band alignment of CdS/ZnSe coupled quantum dots to be of type-II in nature, where the effective gap is smaller than the gap of either  of its components. We have analyzed in details the nature of chemical bonding at the interface, in particular, the calculation of the energy resolved charge density not only clarified the alignment of the bands at the interface but also provided a direct estimate of the band offsets. Our calculations also indicate the important role of the anion-$p$ states in deciding the HOMO offset. The importance of the anion $p$ states was further clarified by considering CdSe/ZnSe coupled quantum dot with a common anion Se and the calculations revealed absence of HOMO offset with a quasi type-II band alignment. We have illustrated that the offsets of HOMO and LUMO can be tuned by changing the sizes of the components of the coupled quantum dot, thereby providing an additional control parameter to tune band gap and optical properties. Our investigations also suggest that formation of alloy near the interface does not change the band offsets substantially but affects the spatial localization of the states. Comparing the influence of strain on coupled quantum dots and core/shell nanowire of CdS/ZnSe, we conclude that the strain at the interface play a crucial role on the electronic structure of core/shell nanowires and hardly affects the electronic structure at the interface of a coupled quantum dot. This is due to the fact that the effective area of interface of the coupled dots is small and as a consequence a small lattice mismatch does not lead to much stress. We have illustrated that quantum confinement primarily controls the properties of coupled quantum dot and should therefore be an ideal candidate for the design of a quantum device.

\section{Acknowledgements}
ID and SA thank Department of Science and Technology, Govt.\ of India for financial support.


%

\end{document}